\documentclass[12pt,preprint]{aastex}

\sfcode`A=1000 \sfcode`B=1000 \sfcode`C=1000 \sfcode`D=1000
\sfcode`E=1000 \sfcode`F=1000 \sfcode`G=1000 \sfcode`H=1000
\sfcode`I=1000 \sfcode`J=1000 \sfcode`K=1000 \sfcode`L=1000
\sfcode`M=1000 \sfcode`N=1000 \sfcode`O=1000 \sfcode`P=1000
\sfcode`Q=1000 \sfcode`R=1000 \sfcode`S=1000 \sfcode`T=1000
\sfcode`U=1000 \sfcode`V=1000 \sfcode`W=1000 \sfcode`X=1000
\sfcode`Y=1000 \sfcode`Z=1000

\newcommand{\js}{}
\newcommand{\etal}{{\js et~al.\/}}
\newcommand{\eg}{{\js e.g.\/}}
\newcommand{\ie}{{\js i.e.\/}}
\newcommand{\Lsol}{L$_{\odot}$}
\newcommand{\Msun}{\mbox{$M_\sun$}}

\newcommand{\sst}{{\it Spitzer Space Telescope}}
\newcommand{\s}{{\it Spitzer}}

\slugcomment{ms 60202v2 short \today}
\shorttitle{M~81}
\shortauthors{Willner et al.}

\begin{document}


\title{IRAC Observations of \object[M 81]{M81}}


\author{
S.\ P.\ Willner,\altaffilmark{1}
M.~L.~N.~Ashby,\altaffilmark{1} 
P.~Barmby,\altaffilmark{1} 
G.~G.~Fazio,\altaffilmark{1} 
M.~Pahre,\altaffilmark{1}
H.~A.~Smith,\altaffilmark{1}
Robert C.\ Kennicutt, Jr.,\altaffilmark{2}
Daniela Calzetti,\altaffilmark{5}
Daniel A.\ Dale,\altaffilmark{3}
B.\ T.\ Draine,\altaffilmark{4}
Michael W.\ Regan,\altaffilmark{5} 
S.\ Malhotra,\altaffilmark{5}
Michele D.\ Thornley,\altaffilmark{6,5}
P.~N.\ Appleton,\altaffilmark{7}
D.~Frayer,\altaffilmark{7} 
G.~Helou,\altaffilmark{7}
S.~Stolovy\altaffilmark{7}
L.~Storrie-Lombardi,\altaffilmark{7}
} 
\altaffiltext{1}{Harvard-Smithsonian Center for Astrophysics, 60 Garden Street,
Cambridge, MA02138}
\altaffiltext{2}{Steward Observatory, University of Arizona, Tucson, AZ 85721}
\altaffiltext{3}{Department of Physics \& Astronomy, University of Wyoming,
Laramie, WY 82071}
\altaffiltext{4}{Princeton University Observatory, Princeton, NJ 08544}
\altaffiltext{5}{Space Telescope Science Institute, 3700 San Martin Drive,
Baltimore, MD 21218}
\altaffiltext{6}{Department of Physics, Bucknell University,
Lewisburg, PA 17837}  
\altaffiltext{7}{Spitzer Science Center, Caltech, 1200 E. California
Blvd., Pasadena, CA 91125}


\begin{abstract}
IRAC images of M81 show three distinct morphological constituents: a
smooth distribution of evolved stars with bulge, disk, and spiral arm
components; a clumpy distribution of dust emission tracing the spiral
arms; and a pointlike nuclear source.  The bulge stellar colors are
consistent with M-type giants, and the disk colors are consistent
with a slightly younger population.    The dust emission generally
follows the blue and ultraviolet emission, but there are large areas
that have dust emission without ultraviolet and smaller areas with
ultraviolet but little dust emission.  The former are presumably
caused by extinction, and the latter may be due to cavities in the
gas and dust created by supernova explosions.  The nucleus appears
fainter at 8~\micron\ than expected from ground-based 10~\micron\
observations made four years ago.
\end{abstract}






\section{Introduction}

Spiral galaxies are valuable laboratories for stellar evolution, stellar
population, and star formation studies.  Detailed observations of the
nearest galaxies are especially valuable, because they allow us to
characterize the integrated properties of the stellar and
interstellar components and their global relationships, while
maintaining sufficient spatial resolution and sensitivity to study
the properties of individual regions.

Infrared studies of spiral galaxies yield insight into both the evolved
stellar content and star formation.  Because 
red giant stars have their  peak emission in the near infrared, 
this wavelength
range is most directly related to the underlying stellar mass
\citep[e.g.,][]{Kauffmann1998,Bell2001}.  
Emission from the hottest dust grains in star forming regions is
detectable at wavelengths longer than about 3~\micron, but grain
emission only becomes a significant fraction of the overall emission
at $\lambda\ga 5$~\micron\ with the rise
of the ``aromatic features'' \citep{helou00,lu03}.
These features are usually attributed to vibrational modes of 
benzene-ring-based materials often called polycyclic 
aromatic hydrocarbons \citep[PAH---][]{leger84}.  This emission
is dominated by transient heating from single-photon events,
so there is no temperature information in the emission colors.

\sst\ observations of spiral galaxies offer a tremendous advance over
previous capabilities.  The vast improvement in sensitivity gives the
ability to map entire galaxies quickly, and the wavelength coverage
allows direct tracing of the stellar and non-stellar components (\eg,
\citealt{pahre04}, who discuss classification schemes based on these
components).
\s\ also provides sufficient angular resolution
to map individual clouds and star forming
regions in nearby galaxies; the IRAC resolution of  $\le2''$
corresponds to a linear scale of $\le$40~pc at a distance of 4 Mpc.

This paper reports initial observations of M81 with the Infrared
Array Camera \citep[IRAC---][]{irac} on \s.  \citet{Gordon} report
companion observations at longer wavelengths, and \citet{helou}
report similar observations of NGC~300.  M81 is one of the
nearest large spiral galaxies at a distance of 3.63~Mpc \citep{K02}.
Its Hubble type is Sab
\citep{RC3}, and its far infrared luminosity is $L_{FIR} = 3 \times
10^9$~\Lsol\ \citep{bgc}.  M81 is a member of a group with $\sim$25
companions and has recently interacted with M82 and NGC~3077
\citep{Y00} and perhaps NGC~2976 \citep{Appleton1988}.  With
isophotal major and minor axis diameters of 27$'$$\times$14$'$, M81
is an excellent target for initial \s\ observations.  M81 also is the
largest galaxy in the Spitzer Infrared Nearby Galaxies Survey
\citep[SINGS---][]{sings}, and these early observations provide a
demonstration of the scientific potential of SINGS for its full
75-galaxy sample.

\section{Observations}

IRAC observations of M81 were made on 2003 November 6.  
Each of 23 positions was observed with six
12-s frames: two repeats at each of three dither positions.
Positions near the center
were observed last to minimize the effects of residual images.  The
visible disk of the galaxy was observed by both of IRAC's
side-by-side fields of view, but the extended sky coverage is toward
the northeast for 3.6 and 5.8~\micron\ and toward the southwest for
4.5 and 8.0~\micron.

The data were processed with Version 8.5.0 of the SSC online
pipeline, using calibration files (`skyflats' and `skydarks') derived
from in-flight data. 
Further processing was carried out in
{\sc iraf}: `skydarks' constructed by median-combining data well outside
the galaxy were subtracted from 3.6 and 5.8$\mu$m frames, a
similarly-constructed `skyflat' was divided into the 8.0 $\mu$m
frames, and background levels of frames in all 4 bands were set to
approximately zero by subtracting the median sky levels. The SSC
``background matching'' software was used to refine this
procedure. Image artifacts\footnote{
Artifacts include banding, muxbleed, and column pulldown and
are described in the IRAC Data Handbook at 
http://ssc.spitzer.caltech.edu/irac/dh/ .}
were corrected (for
cosmetic purposes) with the {\sc iraf imedit} task.  Mosaics combining all
the data were made with the 2004 January 17 version of the SSC
software {\sc mopex}. Mosaics for all four bands were made on the same sky
grid using 0\farcs86 pixels. Because cosmic rays have different
appearances in the four  bands, 
different {\sc mopex} settings were used to optimize cosmic ray
rejection. Figure~1 shows the four resulting images.\footnote{Data
files are available via http://cfa-www.harvard/edu/irac/publications~.}

\section{Analysis}

Inspection of Figure~1 shows three distinct morphological components:
a smooth distribution of starlight prominent in the 3.6 and
4.5~\micron\ images, a clumpy distribution of dust emission seen at
5.8 and 8~\micron, and a pointlike source at the nucleus seen at
4.5~\micron\ and longer wavelengths.  These components are discussed
in turn below.

Quantitative analysis was based on fitting elliptical isophotes
\citep{pahre04} to all four images starting at a radius of 5$''$.
The isophotes as a function of semi-major axis were then fit to disk
$+$ bulge models.
Similar models  were also fit to circular aperture magnitudes, 
and the differences are taken to
indicate the uncertainties.  Observed quantities have no extinction
correction; Galactic extinction is negligible \citep{Schlegel}, and
extinction internal to M81 is patchy and uncertain.\footnote{
For example, \citet{Kaufman1989} find $A_V > 8$ for some of
the narrow dust lanes, but extinction is much smaller over most of the
disk. $A_V=8$ corresponds to $E([3.6]-[4.5]) \approx 0.09$, which
is near the limit of detection in the current data.}

\subsection{Stellar Content}

At 3.6 and 4.5~\micron, the isophotes are well fit by the disk $+$
bulge models as shown in Figure~2.  Parameters are given in Table~1.
The fitting also determines total magnitudes, which are 3.52 and 3.61
(relative to Vega $= 0$) at isophotal wavelengths of 3.54 and
4.50~\micron\ \citep{irac}.  Uncertainties are limited by the \s\
calibration uncertainty of 10\%.
Comparison with the 2MASS $K_S$ total magnitude 3.83 \citep{2mass} shows
that any non-stellar contribution at 3.6~\micron\ is small.  The
derived color $K_S - [3.6] = 0.31$ is the same within the
uncertainties as $K-L = 0.25\pm0.05$ for normal galaxies
\citep{Willner1984}.  
Likewise,  $[3.6] - [4.5] = -0.09$ is very
close to the expected $-0.15$ for an M0~III star (M.~Cohen, private
communication), a spectral type representative of old
stellar populations  \citep{frogel}.\footnote{
The negative, \ie, blue, color for  M stars is
because of  CO absorption in  M star atmospheres in the
4.5~\micron\ band.  K~stars have weaker CO absorption and are
therefore redder than M~stars in the $[3.6]-[4.5]$ color.}

Despite the overall good fit of the [3.6]$-$[4.5] color to late type
stellar colors over most of the galaxy, there are regions that show
redder colors. One region is the nucleus, discussed below.
In addition, the bulge and disk show different overall colors:
$[3.6] - [4.5] = -0.17$ for the bulge and $-0.04$ for the disk.
All explanations for this color difference point to a disk generally
younger than the bulge.  A stellar population even slightly younger ---
dominated by K~stars instead of M~stars --- would have
$[3.6] - [4.5]$ close to zero.  A greater proportion of
dusty interstellar matter in the disk may also cause reddening or, if
the dust is heated, show up in emission, but dust is likely to be a
small contributor to the color difference.\footnote{
If the color difference between disk and bulge is entirely due to
emission from transient-heated hot dust grains in the disk, and the dust
emission follows the $f_\nu \propto \nu^{+0.65}$ spectrum found by
\citet{helou00}, the dust contribution to the 3.6~\micron\ emission
would be 7\% of the disk flux and 4\% of the total flux.  The
\citet{Li2001} dust model predicts a steeper slope, which would make
the 3.6~\micron\ dust contribution 13\% of the disk emission at that
wavelength.} 
There is also a small color gradient
within the bulge.  In contrast to the disk fit, which shows the same
effective radius at 3.6 and 4.5~\micron, the bulge fit shows a larger
effective radius at 3.6 than at 4.5~\micron.  This means the inner
regions of the bulge are  redder than the outer
regions and may indicate a younger stellar component near the
nucleus.

The arm/interarm surface brightness ratio is about 0.5~mag at
3.6~\micron, comparable to values seen in other galaxies at
2.2~\micron\ \citep[\eg][]{Rix1993,Rix1995,Grauer}.  This ratio should be a
good measure of the mass density enhancement in the arms.\footnote{
If short-lived red supergiants contribute significantly to the light
in the spiral arms but little to the interarm regions, the
arm/interarm mass ratio would be smaller than indicated by the
relative surface brightnesses.  However, \citet{Rix1993} found for
M51 that red supergiants are significant only in a single small patch
and not in the overall emission.}
Nearly the same arm/interarm ratio is found from
a blue image \citep{ha} when the exact same areas of the galaxy are
compared.  This agrees with ratios found by \citet{Elmegreen1989}.
In fact, the 3.6~\micron\ contrast is smaller inside a $5'$
radius and larger outside, just like the contrast in blue light.
However, the qualitative appearance in blue light differs from that
at 3.6~\micron; the latter shows a smoother distribution, being less
affected by star clusters or extinction patches.

\subsection{Dust Content}

One of the most striking results from the IRAC images is the complex
structure of the non-stellar emission, as traced in the 5.8 and
8~\micron\ images. Based on the morphology and colors, we assume that
the 3.6~\micron\ image is a direct measure of the stellar light.  We
have therefore subtracted the 3.6~\micron\ image, scaled according to
the expected color of an M0 star \citep{pahre04}, from the 5.8 and
8~\micron\ images to produce images of non-stellar emission.  
The 8~\micron\ non-stellar image is shown in Figure~3.
Non-stellar flux densities are 2.3 and 5.9~Jy at 5.8 and 8.0~\micron\
respectively, 35\% and 69\% of the emission at these wavelengths.

Qualitatively the non-stellar emission traces the same spiral
structure as other gas tracers such as the H~I
\citep[cf.\null][]{Rots1975,Appleton1981,adler96} and optical dust
lanes, as well as young stellar tracers including H$\alpha$,
ultraviolet continuum emission \citep{hut}, and the radio continuum
(Kaufman et al.\ 1989).  When the structure of the IR emission is
examined in more detail, however, it is most strongly correlated with
the optical dust lanes (as might be expected).  Figure~3 shows large
regions where there is strong emission at 8~\micron\ but no
detectable ultraviolet emission.  Detectable dust emission extends to
the outer reaches of the star forming disk but not as far as the
outer H~I spiral arms \citep{adler96}, showing a decline in the
product of PAH abundance and intensity of the stellar radiation.

The non-stellar emission is both more spatially extended than the
stellar distribution (as shown in Figure~2(d) by the rising ratio of
non-stellar to stellar emission beyond $100''$) and much clumpier (as
shown in Figure~1). The arm/interarm ratio is 3 in the 8~\micron\
image (including starlight), similar to the ratio for NGC~6946 at 7
and 15~\micron\  \citep{Malhotra1996}.  The non-stellar image (\ie,
with starlight subtracted) is so clumpy that an ``arm surface
brightness'' is virtually meaningless, but even modest clumps have
surface brightnesses an order of magnitude greater than interarm
regions,\footnote{ 
The interarm regions probably have non-zero
surface brightness, but the uncertainties are large and hard to
quantify because the result depends so critically on what stellar
spectrum is subtracted and on the still poorly-known IRAC
calibration for extended sources.}  and the brightest clumps are more
than 3 times brighter still.


The interstellar medium emission at the 5.8 and 8~\micron\ is likely
dominated by aromatic features usually attributed to PAH, as
demonstrated for instance by the ISO-PHT spectra of galaxies analyzed
by \citet{lu03}.  \citet{Rig99} observed the central region of M81
($24''\times24''$ beam) with ISOPHOT-S and detected emission in the
7.7~\micron\ PAH feature but found a small feature/continuum ratio of
$<$0.115.  The low ratio compared to other galaxies may result from
the large angular size of M81 relative to the
more distant galaxies observed and thus the
lesser disk contribution in  the same fixed beam size.  Both
starlight from the bulge and AGN
emission (Section~3.3) in the M81 nucleus will dilute the
aromatic feature emission, which mostly comes from the disk (Figure~3). 
Aromatic feature emission measured with ISO-CAM
in the 6.75~\micron\ filter correlates well with
H-alpha emission \citep{roussel01} (although the
H$\alpha$ to 15~\micron\ correlation may be
stronger---\citealt{Sauvage1996}).  PAH at 6.75~\micron\ was 
therefore proposed as a reliable star formation estimator for typical
galaxy disks. However, old, cool stars
can also excite aromatic feature emission \citep{Li2002}, and such emission is
therefore not expected to be a perfect star formation indicator, any
more than far infrared luminosity or any other indicator.

Figure~3 shows that
the non-stellar dust emission overall tracks the near ultraviolet
emission (NUV), but there are both regions with dust emission and no
NUV and with NUV but no dust emission.  PAH without NUV can be
explained by dust extinction; one magnitude of visual extinction,
which would heavily attenuate the NUV, corresponds to an
undetectable 0.01~mag of differential extinction between 3.6 and
4.5~\micron.  Seeing NUV light but not dust emission is harder to explain,
but perhaps cavities cleared by supernova explosions are
responsible.  A signature of such cavities is clumpy H$\alpha$
\citep{ha} and
8~\micron\ emission surrounding an area dark at these wavelengths but
bright in NUV.  This morphology can be explained by a supernova
explosion having cleared gas and dust from a young star cluster,
leaving behind and revealing the hot stars.

\subsection{Nucleus}

The nucleus of M81 has been classified as a {\sc liner}
\citep{heckman} or Seyfert~1 \citep{peimbert} with an estimated black
hole mass $7\times10^{7}$~\Msun\ \citep{devereux}.  Like many
galactic nuclei with inferred black holes, the M81 nucleus seems
relatively inactive.  Nuclear
emission varies by a factor of about two in visible light
\citep{devereux} and near 10~\micron\ \citep{grossan} on time scales
estimated at about 25~years.  The ground-based 10~\micron\ data at
$\sim$arcsec resolution suggest a pointlike nucleus.

The new IRAC data suggest that the nuclear source has varied in the
previous four years. IRAC flux densities in a $3\farcs5$ square aperture ({\em
including} starlight) are 109, 75, 65, and 56~mJy at 3.6 to
8~\micron\ respectively with 
uncertainties of about 10\%.\footnote{
These flux densities are relative to Vega flux densities of 277.5,
179.5, 116.6, and 63.1~Jy and are
{\em without} subtracting any stellar
component.  They include corrections for IRAC aperture effects and
are thus directly comparable to ground-based 
observations made with a large chopper throw.  Chopping with a small
throw (\eg, $9\farcs6$ as used by Grossan \etal) would give smaller
flux densities than with a large throw.} 
The 8~\micron\ flux density can be compared with the 1999 10~\micron\
flux density of 159~mJy \citep{grossan}.  
The spectral shape of the nuclear emission between 8 and 10~\micron\
is not known, although the ISO spectrum in a $24''$ beam shows a steep
rise \citep{Rig99}.  Nevertheless, it seems hard to account for the
difference in measured flux densities (159~mJy at 10~\micron\ in
1999; 56~mJy at 8~\micron\ in 2003) by spectral shape alone.  
If the difference is due to variability,
it will strongly support the idea that we are seeing an AGN component.

In order to examine the nuclear morphology, the bulge $+$ disk model
fits were subtracted from the IRAC images at each wavelength.  The
results are shown in Figure~3. A residual pointlike source is quite
evident at 4.5~\micron\ and longer wavelengths and might be present
at 3.6~\micron\ as well, although the brightness of the stellar bulge
at 3.6~\micron\ and nearby negative values in the difference image
make it hard to be sure.  Flux densities of the apparent pointlike
source at the four IRAC wavelengths are 6, 21, 22, and 39~mJy
respectively, with the first being uncertain by more than a factor of
two.  \citet{Davidge1999} found an unresolved source at the
M81 nucleus on the basis of J-K color, and \citet{Quillen2001}
measured a flux density for the point source of 13~mJy at
1.6~\micron.  Considering the uncertainties and possible variability,
the spectral energy distribution from 1.6 to 8~\micron\ might be
consistent with either a power law ($F_\nu \propto \nu^{-0.7}$) or
emission from dust having a temperature range of 400--1000~K and
emissivity that decreases with wavelength.

The images also show a filament or bar of material leading from a
ringlike structure at $\sim$1~kpc ($1'$) to an inner disk and then to
the AGN.  The prominence of the ring at 8~\micron\ suggests that much
of its emission is from dust and probably mostly aromatic features.  These
inner features are probably related to the structures seen in HST
H$\alpha$ images \citep{devereux}.

\section{Conclusions}

\begin{enumerate}
\item
IRAC images of M81 delineate three components by both morphology and
energy distribution: evolved stars organized in bulge and disk
components; a clumpy, dusty interstellar medium with star forming
regions; and a pointlike active nucleus.

\item
The disk is redder than the bulge in the $[3.6]-[4.5]$ color.  This
could be explained by reddening in the disk, by small amounts of dust
emission in the disk, or by a younger stellar population in the disk.

\item
The arm/interarm surface brightness contrast at 3.6~\micron\ is
0.5~magnitudes, in agreement with previous results
in blue light.  For normal stellar populations, this value should
represent the stellar mass density contrast.

\item
The nucleus of M81 shows an apparent point source.  Its flux density
appears to have decreased by something like a factor of 3 in the past
four years, and its SED can be represented by a non-thermal power law
or by hot dust emission.

\item
Star-forming regions are identified by their dust emission, probably
mostly in the aromatic feature bands.  Overall the dust emission
traces the spiral arms as seen in visible and near ultraviolet light,
but there are many regions where dust emission is present and NUV is
absent and some where the reverse is true.  The ubiquity of dust
emission implies that absorption affects both the
distribution and amount of NUV light seen from star-forming regions.

\end{enumerate}




\acknowledgments

This work is based on observations made with the \sst, which is
operated by the Jet Propulsion Laboratory, California Institute of
Technology under NASA contract 1407. Support for this work was
provided by NASA through contract  1256790 issued by JPL/Caltech.
Support for the IRAC instrument was provided by NASA through contract 
960541 issued by JPL.  M.~A.~P. acknowledges NASA/LTSA grant
NAG5-10777. 

IRAF is distributed by the National Optical Astronomy Observatories,
which are operated by the Association of Universities for Research
in Astronomy, Inc., under cooperative agreement with the National
Science Foundation.

This research has made use of the NASA/IPAC Extragalactic Database
(NED), which is operated by the Jet Propulsion Laboratory, California
Institute of Technology, under contract with the National Aeronautics
and Space Administration.



\noindent
Facilities: \facility{Spitzer(IRAC)}\\
Dataset: \dataset{ads/sa.spitzer\#0006629120}






\clearpage

\begin{figure}
\epsscale{.80}
\plotone{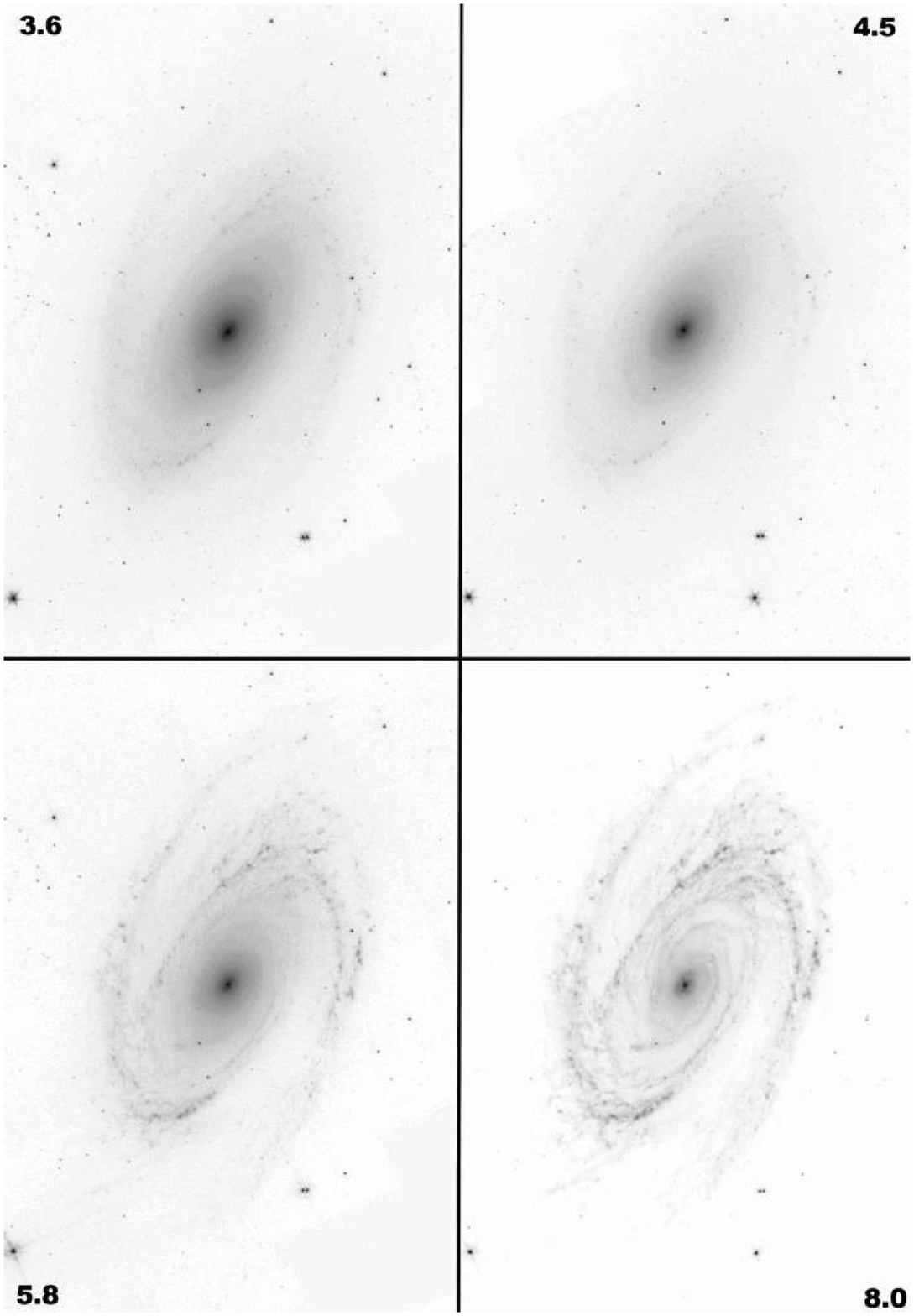}
\caption{IRAC images in the four bands.  North is up and east to the
left in all panels.  Grey scale is $\sqrt{\log}$ to show faint
structure.}
\end{figure}

\clearpage
\begin{figure}
\plotone{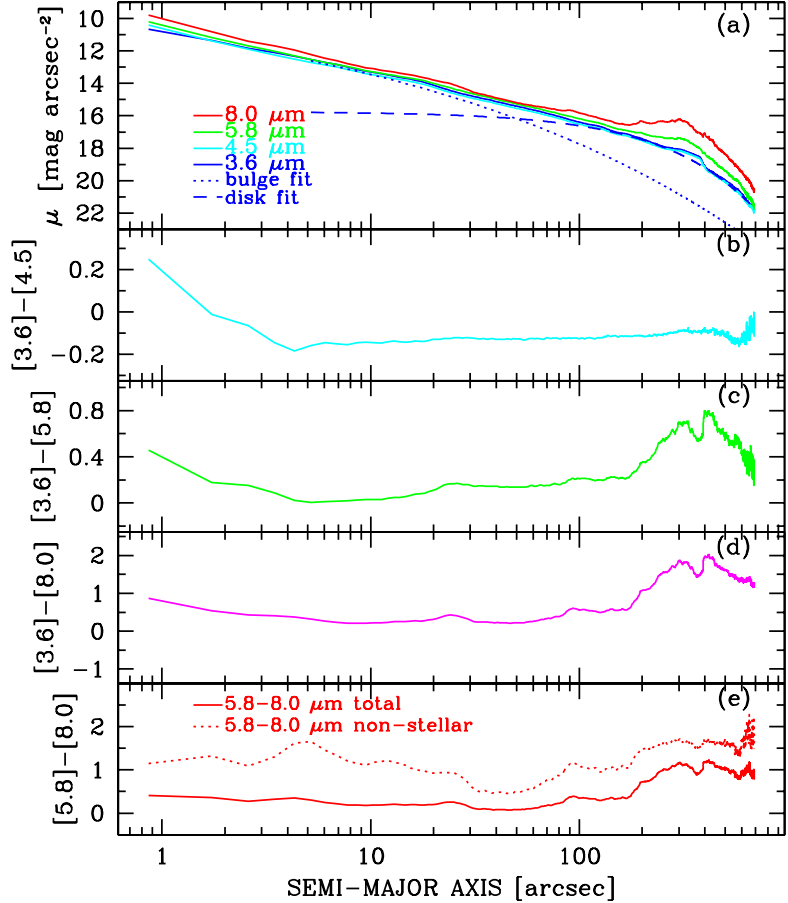}
\caption{Surface brightness profiles in elliptical isophotes.  Solid
lines in panel (a) show the measured surface brightness profiles for
all four IRAC bands.  Dashed lines show the bulge $+$ disk model for
3.6~\micron.  Panels (b), (c), and (d) and the solid line in panel
(e) show the measured surface brightness profiles for three different
IRAC colors.  The dashed line in panel (e) shows the surface
brightness profile for the non-stellar emission alone (\ie, after
starlight has been subtracted) in the $[5.8]-[8.0]$ color.}
\end{figure}

\begin{figure}
\epsscale{.70}
\plotone{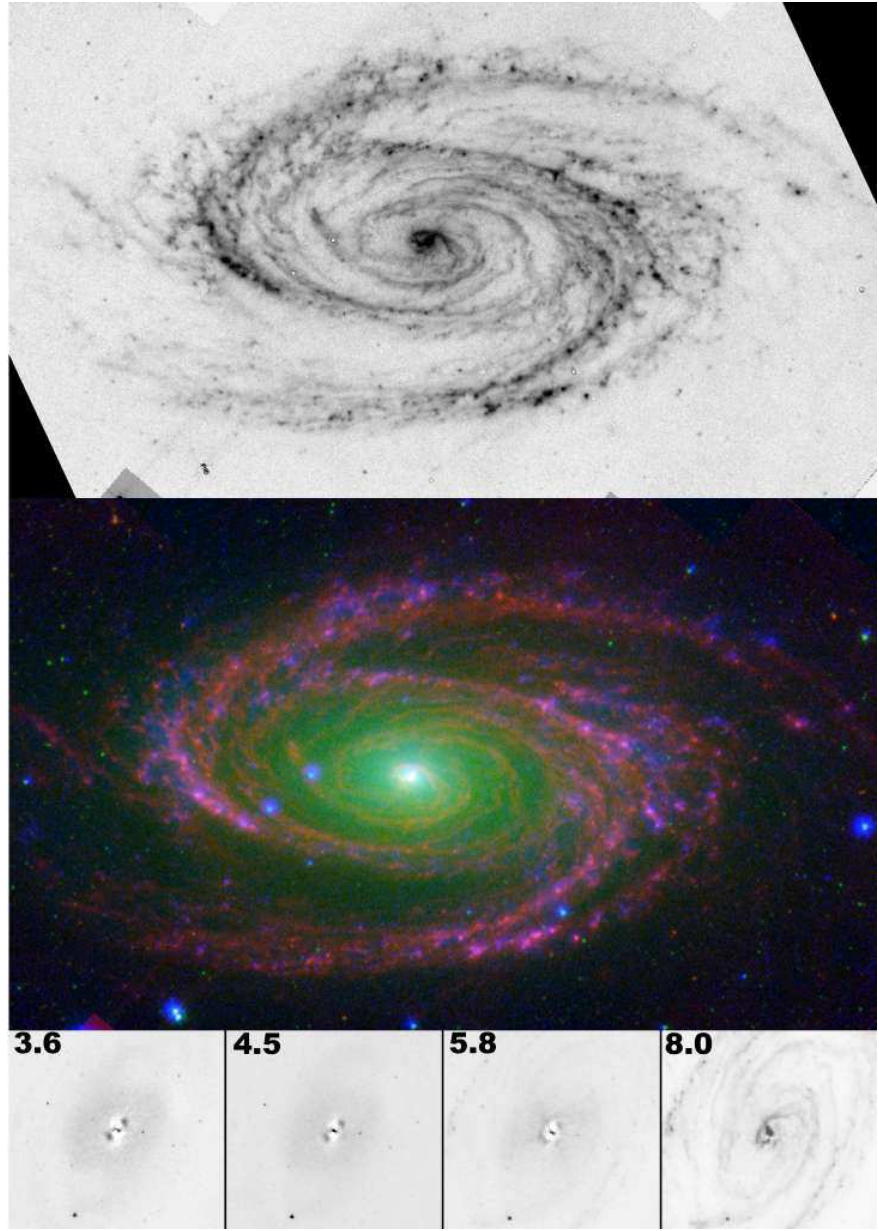}
\caption{Upper panel: 8~\micron\ non-stellar radiation.  Middle panel:
composite color image with blue from the near ultraviolet \citep{hut}, green
from the 3.6~\micron\ image, and red the non-stellar emission from the
upper panel.  North is $65^\circ$ clockwise from the top, and the
images are 20$'$ horizontally.  Three regions showing NUV but not
8~\micron\ emission are visible in the outer spiral arm above the
nucleus.  Lower panels: 
nuclear regions after subtraction of the bulge $+$ disk model.
Images are $5'= 5.3$~kpc square, and North is up.
Table~1 gives model parameters for 3.6~\micron; values for other bands are
similar.  White regions near the nucleus represent negative values,
i.e., areas where the model surface brightness exceeds the observed
surface brightness.}
\end{figure}

\clearpage

\clearpage






\begin{deluxetable}{lcc}
\tabletypesize{\normalsize}
\tablewidth{0pt}
\tablecaption{M81 Stellar Properties  
\label{tbl1}}
\tablehead{
\colhead{Parameter} &
\colhead{3.6~\micron\tablenotemark{a}}&
\colhead{$B$\tablenotemark{b}}}
\startdata
Bulge total magnitude  &4.62$\pm$0.05 & 8.99 \\
Bulge half-light semi-major axis $a_B$ (arcsec) & 54 & 85 \\
Bulge surface brightness\tablenotemark{c} 
  (mag~arcsec$^{-2}$) &14.86$\pm$0.03 & 20.29 \\
Bulge ellipticity ($1-b_B/a_B$) & 0.30 & 0.26 \\
Disk total magnitude   &4.01$\pm$0.05 & 8.23  \\
Disk half-light semi-major axis $a_D$ (arcsec) & 242 & 288 \\
Disk  surface brightness\tablenotemark{c} 
  (mag~arcsec$^{-2}$) &17.09$\pm$0.23 & 21.71\\
Disk ellipticity ($1-b_D/a_D$) & 0.53 & 0.53 \\
Galaxy total magnitude &3.52$\pm$0.02 & 7.79  \\
Bulge to disk ratio    &0.57$\pm$0.05 & 0.49 \\
\enddata
\tablenotetext{a}{Indicated uncertainties are those from the fitting
procedure; they do not include overall calibration uncertainties of
approximately 10\%.}
\tablenotetext{b}{Parameters from \citet{Moellenhoff2004}.  Bulge fit at $B$
is based on an $r^{-0.29}$ profile instead of $r^{-0.25}$ as used for
3.6~\micron.}
\tablenotetext{c}{Mean surface brightness for area within the
half-light ellipse.}


\end{deluxetable}





\begin{thebibliography}{}


\bibitem[Adler \& Westpfahl(1996)]{adler96}
Adler, D.~S., \& Westpfahl, D.~J. 1996, \aj, 111, 735

\bibitem[Appleton, Davies, \& Stephenson(1981)]{Appleton1981} 
Appleton, P.~N., Davies, R.~D., \& Stephenson, R.~J.\ 1981, \mnras, 195, 
327 

\bibitem[Appleton \& van der Hulst(1988)]{Appleton1988} 
Appleton, P.~N.~\& van der Hulst, J.~M.\ 1988, \mnras, 234, 957 

\bibitem[Bell \& de Jong(2001)]{Bell2001} 
Bell, E.~F.~\& de Jong, R.~S.\ 2001, \apj, 550, 212 


\bibitem[Cheng \etal(1997)]{ha}
Cheng, K.-P., Collins, N., Angione, R., Talbert, F., Hintzen, P., 
Smith, E.P., Stecher, T., and the UIT Team 1997, {\it UV/Visible Sky
Gallery}, CD-ROM, data available from NASA Extragalactic Database (NED)
    

\bibitem[Davidge \& Courteau(1999)]{Davidge1999} 
Davidge, T.~J.~\& Courteau, S.\ 1999, \aj, 117, 2781

\bibitem[de Vaucouleurs et al.(1991)]{RC3} 
de Vaucouleurs, G., de Vaucouleurs, A., Corwin, H.~G., Buta, R.~J., 
Paturel, G., \& Fouque, P.\ 1991, Volume 1-3, XII, 2069 pp.~7 figs..~ 
Springer-Verlag Berlin Heidelberg New York  

\bibitem[Devereux \etal(2003)]{devereux} 
Devereux, N., Ford, H., Tsvetanov, Z., \& Jacoby, G.\ 2003, \aj, 125, 1226 

\bibitem[Elmegreen, Seiden, \& Elmegreen(1989)]{Elmegreen1989} 
Elmegreen, B.~G., Seiden, P.~E., \& Elmegreen, D.~M.\ 1989, \apj, 343, 602

\bibitem[Fazio \etal(2004)]{irac}
Fazio, G.~G.\ \etal\ 2004, \apjs, this volume.

\bibitem[Frogel(1985)]{frogel} 
Frogel, J.~A.\ 1985, \apj, 298, 528 

\bibitem[Gordon \etal(2004)]{Gordon}
Gordon, K.\ D.\ \etal, 2004, \apjs, this volume

\bibitem[Grauer \& Rieke(1998)]{Grauer}
Brauer, A.\ D.\ \& Rieke, M.\ J.\ 1998, \apjs, 116, 29.

\bibitem[Grossan, Gorjian, Werner, \& Ressler(2001)]{grossan}
Grossan, B., Gorjian, V., Werner, M., \& Ressler, M.\ 2001, \apj,
563, 687 

\bibitem[Heckman(1980)]{heckman} 
Heckman, T.~M.\ 1980, \aap, 87, 152 

\bibitem[Helou \etal(2004)]{helou}
G.\ Helou \etal\ 2004, \apjs, this volume.

\bibitem[Helou \etal(2000)]{helou00}
G.\ Helou, N.~Y.\ Lu, M.~W.\ Werner, S.\ Malhotra \& N.~A.\ Silbermann 2000,
\apjl, 532, L21  


\bibitem[Jarrett et al.(2003)]{2mass}
Jarrett, T.~H., Chester, T., Cutri, R., Schneider, S.~E., \& Huchra,
J.~P.\ 2003, \aj, 125, 525

\bibitem[Karachentsev et al.(2002)]{K02}
Karachentsev, I.~D.\ et al.\ 2002, \aap, 383, 125

\bibitem[Kauffmann \& Charlot(1998)]{Kauffmann1998} 
Kauffmann, G.~\& Charlot, S.\ 1998, \mnras, 297, L23 

\bibitem[Kaufman et~al.(1989a)]{kaufman89}
Kaufman, M. et~al. 1989a, \apj, 345, 674

\bibitem[Kaufman, Elmegreen, \& Bash(1989b)]{Kaufman1989} 
Kaufman, M., Elmegreen, D.~M., \& Bash, F.~N.\ 1989b, \apj, 345, 697 

\bibitem[Kennicutt et~al.(2003)]{sings}
Kennicutt, R.~C. et~al.\ 2003, \pasp, 115, 928 

\bibitem[Leger \& Puget(1984)]{leger84}
Leger, A.\ \& Puget, J.-L., 1984, \aap, 137, L5.

\bibitem[Li \& Draine(2001)]{Li2001} 
Li, A.~\& Draine, B.~T.\ 2001, \apj, 554, 778 

\bibitem[Li \& Draine(2002)]{Li2002} 
Li, A.~\& Draine, B.~T.\ 2002, \apj, 572, 232 


\bibitem[Lu \etal(2003)]{lu03}
Lu, N. et al. 2003, \apj, 588, 199

\bibitem[Malhotra et al.(1996)]{Malhotra1996} 
Malhotra, S., et al.\ 1996, \aap, 315, L161 

\bibitem[Marcum et al.(2001)]{hut}
Marcum, P.~M., et al.\ 2001, \apjs, 132, 129 

\bibitem[M{\" o}llenhoff(2004)]{Moellenhoff2004} 
M{\" o}llenhoff, C.\ 2004, \aap, 415, 63

\bibitem[Pahre \etal(2004)]{pahre04}
Pahre, M.~A., Ashby, M.~L.~N., Fazio, G.~G., \& Willner, S.~P.\ 2004,
\apjs, this volume


\bibitem[Peimbert \& Torres-Peimbert(1981)]{peimbert} 
Peimbert, M.~\& Torres-Peimbert, S.\ 1981, \apj, 245, 845 

\bibitem[Quillen et al.(2001)]{Quillen2001} 
Quillen, A.~C., McDonald, C., Alonso-Herrero, A., Lee, A., Shaked, S., 
Rieke, M.~J., \& Rieke, G.~H.\ 2001, \apj, 547, 129



\bibitem[Rice et al.(1988)]{bgc}
Rice, W., Lonsdale, C.~J., Soifer, B.~T., Neugebauer, G., Koplan,
E.~L., 
Lloyd, L.~A., de Jong, T., \& Habing, H.~J.\ 1988, \apjs, 68, 91 


\bibitem[Rigopoulou et al.(1999)]{Rig99}
Rigopoulou, D., Spoon, H.~W.~W., Genzel, R., Lutz, D., Moorwood,
A.~F.~M., \& Tran, Q.~D.\ 1999, \aj, 118, 2625

\bibitem[Rix \& Rieke(1993)]{Rix1993} 
Rix, H.~\& Rieke, M.~J.\ 1993, \apj, 418, 123 

\bibitem[Rix \& Zaritsky(1995)]{Rix1995} 
Rix, H.~\& Zaritsky, D.\ 1995, \apj, 447, 82

\bibitem[Rots \& Shane(1975)]{Rots1975} 
Rots, A.~H.~\& Shane, W.~W.\ 1975, \aap, 45, 25 

\bibitem[Roussel \etal(2001)]{roussel01}
Roussel, H., Sauvage, M., Vigroux, L. \& Bosma, A.\ 2001, \aap, 372, 427


\bibitem[Sauvage et al.(1996)]{Sauvage1996} 
Sauvage, M., et al.\ 1996, \aap, 315, L89 

\bibitem[Schlegel, Finkbeiner, \& Davis(1998)]{Schlegel}
Schlegel, D.~J., Finkbeiner, D.~P., \& Davis, M.\ 1998, \apj, 500,
525 


\bibitem[Willner et al.(1984)]{Willner1984}
Willner, S.~P., Fabbiano, G., Elvis, M., Ward, M., Longmore, A., \& 
Lawrence, A.\ 1984, \pasp, 96, 143 

\bibitem[Yun, Ho, \& Lo(2000)]{Y00} Yun, M.~S., Ho, 
P.~T.~P., \& Lo, K.~Y.\ 2000, ASP Conf.~Ser.~217: Imaging at Radio through 
Submillimeter Wavelengths, 374 



\end{thebibliography}
\end{document}